\documentclass[%
notitlepage,
twocolumn,
amsmath,amssymb,fleqn,
prb,
floatfix
]{revtex4-2}

 \usepackage{graphicx}  
  \usepackage{xcolor}  \usepackage[colorlinks,citecolor=blue,linkcolor=blue,urlcolor=blue,bookmarks=false,hypertexnames=true]{hyperref}

\usepackage{graphicx}
\usepackage{stackengine}
\usepackage{dcolumn}
\usepackage{bm}
\usepackage{xcolor}
\usepackage{mathtools}
\usepackage{bbold}
\usepackage{subfigure}
\usepackage{mathtools} 
\usepackage{nicefrac}
\usepackage{pgfplots}
\usepackage{float}

\usepgfplotslibrary{external}
\pgfplotsset{compat=1.18}
\newcommand{\ket}[1]{| #1 \rangle}

\begin{document}
	
\preprint{}

\title{High-Fidelity Entangling Gates for Electron and Nuclear Spin Qubits in Diamond}

\author{Regina Finsterhoelzl}
\email{regina.finsterhoelzl@uni-konstanz.de}
\author{Wolf-R\"udiger Hannes}
\email{wolf-ruediger.hannes@uni-konstanz.de}
\author{Guido Burkard}
\email{guido.burkard@uni-konstanz.de}
\affiliation{University of Konstanz, Department of Physics, 78462 Konstanz, Germany}

\date{\today}

\begin{abstract}
Motivated by the recent experimental progress in exploring the use of a nitrogen-vacancy (NV) center in diamond as a quantum computing platform, we propose schemes for fast and high-fidelity entangling gates on this platform. Using both analytical and numerical calculations, we demonstrate that synchronization effects between resonant and off-resonant transitions may be exploited such that spin-flip errors due to strong driving may be eliminated by adjusting the gate time or the driving field. This allows for fast, high fidelity entangling operations between the electron spin and one or several nuclear spins. We investigate a two-qubit system where the NV center is comprised of a $^{15}$N atom and a qubit-qutrit system for the case of a $^{14}$N atom. In both cases, we predict a complete suppression of off-resonant driving errors for two-qubit gates when addressing the NV electron spin conditioned on states of nuclear spins of the nitrogen atom of the defect. Additionally, we predict fidelities $>0.99$ for multi-qubit gates when including the surrounding $^{13}$C atoms in the diamond lattice in the conditioned logic.
\end{abstract}

%Use showkeys class option if keyword
%display desired
\maketitle

\section{Introduction}
Among the candidates for a solid-state hardware platform for quantum technologies, single defects in solids have attracted increasing attention \cite{Awschalom2018,Zwanenburg2013,Weber2010}. 
Next to tin vacancy centers \cite{Pasini2023} or silicon vacancy centers in diamond \cite{Nguyen2019}, and vacancy centers in silicon carbide \cite{Bourassa2020,Castelletto2020,Tissot2022,Castelletto2024}, the negatively charged nitrogen vacancy center (NV center) in diamond has recently been actively investigated as a platform for quantum sensing, quantum networks, and as a near-term candidate for quantum computation \cite{Wrachtrup2006,Jelezko2006,Doherty2013,Dobrovitski2013,Degen2017,Nemoto2014,Ruf2021,Pezzagna2021,Sun2023,Sekiguchi2023}. 
\begin{figure}[b]
    \centering
    \includegraphics[width=0.4\linewidth]{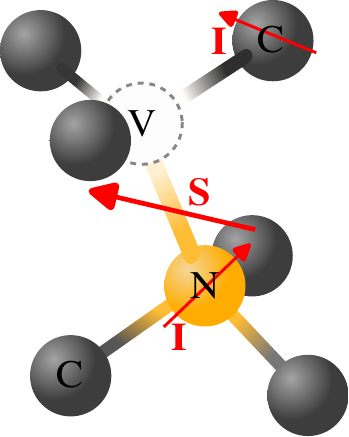}
    \caption{Sketch of a nitrogen-vacancy (NV) center in diamond, consisting of an empty site in a diamond lattice next to a nitrogen atom replacing one carbon atom. In its negative charge state, the NV ground state is a spin triplet, which couples by hyperfine interaction to the intrinsic nitrogen nuclear spin as well as to the nuclear spins of surrounding carbon $^{13}$C atoms in the host diamond lattice.}
    \label{fig:NVcenter}
\end{figure}

The NV center consists of a nitrogen atom neighbouring an empty site in a diamond lattice. Due to the high isolation from its environment, the electron spin of this system has promising properties for its use in quantum information processing applications: It features long coherence times \cite{Taminiau2014,Abobeih2018}, even at elevated temperatures up to room temperature, and is optically addressable for initialization \cite{Aslam2013} and readout \cite{Robledo2011}. 
By hyperfine interaction, the electron spin couples to the intrinsic nuclear nitrogen spin forming a two-qubit (or qubit-qutrit) platform. Additional qubits arise by controlling the coupling to the long-lived nuclear spins of naturally occurring carbon-13 atoms in the host diamond lattice, providing a quantum register for storing and processing quantum states, where full coupling and control of nine nuclear spins has been demonstrated experimentally \cite{Taminiau2012,Kolkowitz2012,Zhao2012_2,Bradley2019}. 
Scaling may be realized by dipolar-mediated electron-electron coupling between neighbouring NV centers \cite{Neumann2010}, and long-range connectivity by cavity-mediated electron-photon coupling \cite{Kubo2010,Auer2016,Burkard2017,Hannes2024} or by magnon-mediated NV-NV entanglement creation \cite{Fukami2021}. 
Qubit control and manipulation is performed with microwave and radio-frequency (rf) laser pulses where high-fidelity single-qubit as well as entangling electron-nuclear spin gates have been demonstrated experimentally, between the electron spin and the nitrogen nuclear spin as well as electron-nuclear gates including the nuclear spin of carbon-13 atoms, based on dynamical decoupling sequences and composite pulse schemes \cite{Childress2006,Dutt2007,Robledo2011,Fuchs2011,Taminiau2012,Kolkowitz2012,Zhao2012_1,Pfaff2013,Dolde2014,Waldherr2014,Taminiau2014,Rong2015,Bradley2019,Xie2023,Takou2022,Takou2023,Mizuno2024}. Recent proposals also include a holonomic two-qubit gate \cite{Shkolnikov2020} and the use of the SMART protocol \cite{Vallabhapurapu2023}.
The coherence properties of these systems \cite{Onizhuk2023}, together with its central spin connectivity, can be exploited for small quantum algorithms and quantum simulation \cite{Ruh2022}, as well as quantum error correcting codes \cite{Waldherr2014,Taminiau2014,Cramer2016,Finsterhoelzl2022,Debone2024} and fault-tolerant operations \cite{Abobeih2022}, further demonstrating the potential of the NV systems as quantum registers. 

One scheme for an entangling gate consists of driving the NV electron spin in a frequency selective manner, conditioned on states of the spins of the surrounding nuclei. Besides the challenge of the central spin decoherence due to the nuclear spin bath \cite{Casanova2016,Maile2023,Chen2023,Bartling2023,Ungar2024}, a main challenge of these gates lies in avoiding the addressing of unwanted transitions in the dense energy spectrum of the hyperfine-coupled system: the hyperfine interaction is always on, and the spectrum is unavoidably crowded. This strongly limits the fidelity of a standard $\pi$-pulse both for single and for multi-qubit gates. This may either be avoided by operating in the weak driving regime \cite{Robledo2011,Dutt2007,Yu2023} at the cost of slower gate operations, or by relying on complex pulse sequences designed digitally by optimal control schemes \cite{Said2009,Dolde2014,Waldherr2014,Rong2015,Xie2023,Kairys2023,Liddy2023}.

In this paper, we follow a different approach. We address the challenge posed by unintended off-resonant driving by proposing a solution which allows for Rabi frequencies beyond the weak driving regime and thus for fast gate operations, while at the same time eliminating the error due to the off-resonant drive completely. 
We show that this is possible  by exploiting synchronization effects between wanted and unwanted transitions, which occur for certain choices of intensity and frequency of the external AC and DC magnetic fields. 
With this, we build on a technique which has emerged in the field of nuclear-magnetic resonance spectroscopy \cite{Koch2022,Glaser2015,Sattler1999,Alexander1961}, and has also been shown theoretically \cite{Russ2018,Heinz2021} and demonstrated experimentally \cite{Noiri2022} for semiconductor spin qubits.
We employ the synchronization technique and tailor it to different types of NV center registers, where we focus on the entangling operation which rotates the central electron spin conditioned on the state of one (C$_n$NOT$_e$) or more (CC$_n$NOT$_e$) nuclear spins in the register. 

The remaining sections of this paper are organized as follows. In Sec.~\ref{sec:model1}, we present the model we use to describe the NV system. Afterwards, in Secs.~\ref{sec1}-\ref{sec3}, we explain the synchronization scheme and its implenmentation for the NV center with one or several coupled nuclei. Here, we first demonstrate the applicability in case of the NV center built with a $^{15}$N nitrogen atom. Next, we investigate the case of a NV-$^{14}$N system. We then extend the analysis to the case of additional coupled $^{13}$C nuclei. Finally, in Sec.~\ref{sec:conclusion}, we summarize our findings and offer perspectives for further research.

\section{Model}
\label{sec:model1}
We consider a single negatively charged NV center, a point defect characterized by a nitrogen atom (either $^{15}$N or, with much higher probability, $^{14}$N), replacing a carbon atom in a diamond lattice where one of the nearest-neighboring sites is vacant (see Fig.~\ref{fig:NVcenter}). Its four orbitals are filled with six electrons leading to an $^3A_2$ electron spin-1 triplet ground state. By the hyperfine interaction, the electron spin couples to the intrinsic nuclear spin of the nitrogen atom leading to a hyperfine splitting of the energy levels. The nitrogen nuclear spin quantum number depends on the isotope and is $I^{15\rm N}=\nicefrac{1}{2}$ ($I^{14\rm N}=1$) in case of a $^{15}N$ ($^{14}$N) atom. The electron spin also couples to the nuclear spins of the surrounding carbon $^{13}$C atoms which may be found with a natural concentration of 1.1\% in the diamond lattice and comprise a nuclear spin $I^{13\rm C}=\nicefrac{1}{2}$. To lift the degeneracy of the electronic spin levels $m_s=\pm 1$ as well as the nuclear spin levels $m_n^{15\rm N}=\pm \nicefrac{1}{2}$ ($m_n^{14\rm N}=\pm 1$) and $m_n^{13\rm C}=\pm \nicefrac{1}{2}$, a static magnetic field $\textbf{B}=(0, 0, B_z)$ is applied along the direction of the principal axis of the NV center. 

The Hamiltonian describing the system is given by
\begin{align}
 H_0=H_e + \sum_j H^j_n + \sum_j H^j_i   ,
 \label{eq:totalhamiltonian}
\end{align}
with the electronic, nuclear, and interaction parts,
\begin{align}
H_e&=DS_z^2+\gamma_{e}B_z  S_z \label{eq:Hamiltonian1},\\
    H^j_n&=Q^j (I^j_z)^2 +\gamma^j_nB_z  I^j_z \label{eq:Hamiltonian2},\\H_i^j&=\textit{\textbf{SA}}^j\textit{\textbf{I}}^j \label{eq:Hamiltonian3},
\end{align}
where we choose energy units such that $\hbar \equiv 1$, $\textit{\textbf{S}}=(S_x,S_y,S_z)$ denotes the electronic spin operator with $S=1$, and where the index $j$ in Equation~\eqref{eq:totalhamiltonian} runs over all coupled nuclear spins. Equation~\eqref{eq:Hamiltonian1} describes the free evolution of the electron spin, where $D/(2\pi)=2.88\,\text{GHz}$ is the magnitude of the zero-field splitting separating the electronic spin levels $m_s=0$ and $m_s=\pm 1$. The reduced electronic gyromagnetic ratio is given by $\gamma_{e} = \mu_B g_e$ with $\mu_B/h=14.00\,\text{GHz/T}$ and $g_e=2.00$. 
Equation~\eqref{eq:Hamiltonian2} describes the free evolution of the nuclear spin with label $j \in \{15\rm N,14\rm N,13\rm C\}$.
This labeling is unique because we are considering up to two nuclear spins which are always of different type.
Here, $Q^j$ denotes the respective quadrupole splitting, $\textit{\textbf{I}}^j=(I^j_x,I^j_y,I^j_z)$ the nuclear spin operator and the reduced nuclear gyromagnetic ratio is given by $\gamma_{n}^j=\mu_N g^j_n$ with the nuclear magneton $\mu_N/h=7.63\,\text{MHz/T}$ and $g_n^j$ the respective nuclear g factor  \cite{Everitt2014}.  Equation~\eqref{eq:Hamiltonian3} captures the hyperfine interaction between the electron and the $j$th nuclear spin described by the hyperfine tensor $\textit{\textbf{A}}^j$.

\begin{figure}
    \centering
\includegraphics[width=\linewidth]{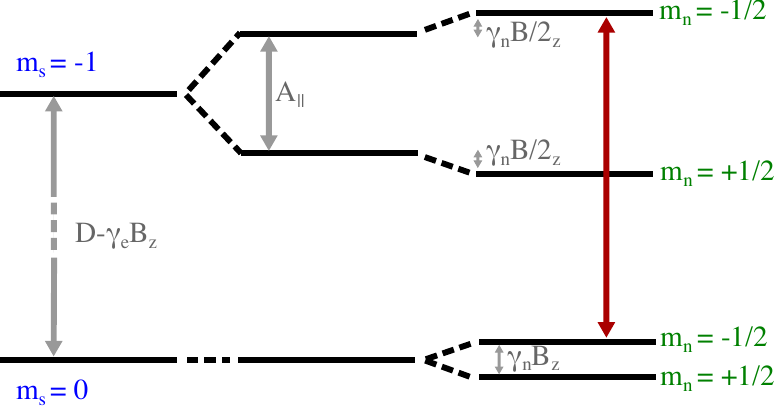}
    \caption{Energy level scheme of an NV center comprising a $^{15}$N atom.  Near the ground-state level anticrossing, the two  electronic spin states $m_s=0$, $m_s=-1$ are relevant at low energy, while the  $m_s=1$ level (not shown) is far detuned. Shown are the eigenenergies of $H_e$ (left), $H_e+H_i^{15\rm N}$ (center), and $H_e+H_i^{15\rm N}+H_n^{15\rm N}$ (right). The red arrow indicates the  transition under study.}
    \label{fig:energylevels15N}
\end{figure}

The NV system can be initialized \cite{Aslam2013,Weber2010} and read out \cite{Robledo2011,Abobeih2022,Abobeih2018,Vorobyov2013} optically via the electron spin and its coupling to the nuclear spins \cite{Cramer2016,Bradley2019,Abobeih2022}. This results in the connectivity of a central-spin system where the electron spin mediates all couplings.
Experimental quantum control has first been demonstrated on the NV electron spin and on strongly coupled nuclear spins, including the nuclear spin of the nitrogen atom as well as the spin of surrounding carbon-13 nuclei where the parallel hyperfine component $A_{||}$ is much larger than the inverse of the coherence time of the electron $T_{2,e}^*$, thus $A_{||} \gg 1/T_{2,e}^*$ \cite{Childress2006,Dutt2007,Robledo2011,Fuchs2011,Pfaff2013,Dolde2014,Waldherr2014,Rong2015}. In this case, hyperfine-split energy levels are resolvable in the optically detected magnetic resonance (ODMR) spectrum, which may be used for conditional operations both on electronic and nuclear qubits. In 2012, three groups independently demonstrated conditional logic with qubits based on weakly coupled carbon-13 atoms using dynamical-decoupling (DD) sequences on the electron spin \cite{Taminiau2012,Kolkowitz2012,Zhao2012_2,Takou2022}, a method which has later been combined with rf-pulses on nuclear spins (DDrf sequence) \cite{Bradley2019}.

\section{Electron-nuclear two-qubit entangling gate}
\label{sec1}

We begin our analysis by deriving a protocol for an NV center including an $^{15}$N atom, thus $j=15N$ in Eq.~\eqref{eq:totalhamiltonian}, Eq.~\eqref{eq:Hamiltonian2} and Eq.~\eqref{eq:Hamiltonian3}. As a spin 1/2, $^{15}$N does not exhibit a quadrupole splitting, $Q^{15N}=0$, while the reduced nuclear gyromagnetic ratio is given by $\gamma_{n}^{15\rm N}/(2\pi)=-4.3\,\text{MHz/T}$ \cite{Everitt2014}.
Due to symmetry properties of the NV center, the nitrogen hyperfine tensor $\textbf{A}^{15\rm N}$ is diagonal both in the optical ground and excited state, with parallel (perpendicular) components $A^{15\rm N}_{||}/(2\pi)=3.03\,\text{MHz}$ ($A^{15\rm N}_{\perp}/(2\pi)=3.65\,\text{MHz}$), thus $H_i^{15\rm N}=A^{15\rm N}_{||}S_z I^{15\rm N}_z+A^{15\rm N}_{\perp}(S_xI^{15\rm N}_x+S_yI^{15\rm N}_y)$. 

The perpendicular hyperfine interaction mixes the electronic and nuclear spin states and is resonant at $B_z^{\rm res}=\frac{D\pm A^{15\rm N}_{||}/2}{\gamma_{e} \mp \gamma_{n}}\approx  102\,\textup{mT}$ in the ground state \cite{Everitt2014}. Typical operation regimes are far detuned from this resonance, i.e. with $B_z \approx 0.5\,T$ where its contribution reduces to a dispersive level shift which is small compared to $D$ and may thus be neglected. Eq.~\eqref{eq:Hamiltonian3} can thus be approximated as
$H_i^{15\rm N} \approx A^{15\rm N}_{||}S_z I^{15\rm N}_z$.
With these considerations, we arrive at the Hamiltonian 
\begin{multline}
        H_0^{15\rm N}=DS_z^2+ \gamma_{e} B_z S_z + \gamma^{15\rm N}_n B_z I^{15\rm N}_z \\
    + A_{||}^{15\rm N}S_z I_z^{15\rm N}. 
    \label{eq:hamiltonian4}
\end{multline}
The resulting energy level scheme is depicted in Fig.~\ref{fig:energylevels15N}.
The two qubits encoded in the system given in Eq.~\eqref{eq:hamiltonian4} are defined with the electron spin states $|m_s=0\rangle \equiv |0\rangle_e$ and $|m_s=-1\rangle \equiv |1\rangle_e$ and the nuclear spin states $|m_n=+\nicefrac{1}{2}\rangle \equiv |0\rangle_n$ and $|m_n=-\nicefrac{1}{2}\rangle \equiv |1\rangle_n$. 
We introduce a driving field polarized along the x-axis with a constant driving strength given by
\begin{equation}
    H_D=\Tilde{B_1} \cos{(\omega_0t)} S_x.
    \label{eq:drive}
\end{equation}
In order to minimize leakage into the $\ket{m_s=+1}$ state of the electron triplet, the strength of the driving field $B_1$ needs to be weak against the detuning difference $\delta = \left|\left|E_{m_s=1}-E_{m_s=0}|-|E_{m_s=-1}-E_{m_s=0}\right|\right|$ \cite{Everitt2014} which is given by
\begin{equation}
    \delta = 
   \begin{cases}
     2\gamma_e B_z, & \text{if} \quad \gamma_e B_z < D,\\
     2D, & \text{if} \quad \gamma_e B_z \geq D.
   \end{cases}
   \label{eq:delta}
\end{equation}

To achieve a nuclear-spin controlled operation on the electron spin C$_n$NOT$_e$, we choose the driving frequency resonant with the $\ket{1}_n\ket{0}_e \leftrightarrow \ket{1}_n\ket{1}_e$ transition, thus $\omega_0=D-B_z\gamma_e+\frac{A_{||}^{15\rm N}}{2}$. We transform the total Hamiltonian $H=H_0+H_D$ into the interaction picture given by the free-evolution (Larmor) frequencies with $U_1=e^{it\gamma^{15\rm N}_n B_z I^{15\rm N}_z}$ and $\Tilde{H}(U_1)=U_1HU_1^\dagger -iU_1\partial_tU_1^\dagger$, and perform a rotating wave approximation in the frame of the frequency of the driving field defined by the unitary transformation $U_2=e^{i\omega_0 tS_z}$ and $\Tilde{H}(U_2)$. With this, the Hamiltonian in the computational subspace with the basis $\ket{00},\ket{01},\ket{10},\ket{11}$ where $\ket{ij}=\ket{i}_n \ket{j}_e $, can be written in the following matrix form,
\begin{equation}
    \Tilde{H}=
    \begin{pmatrix}
        0 &  \nicefrac{B_1}{2} & 0 & 0\\
         \nicefrac{B_1}{2} & -A_{||}^{15\rm N} & 0 & 0 \\
       0  & 0 & 0 & \nicefrac{B_1}{2}\\
        0 & 0  & \nicefrac{B_1}{2} & 0
    \end{pmatrix} ,
    \label{eq:hamiltonian5}
\end{equation}
with $B_1\equiv\Tilde{B_1}/\sqrt{2}$. The dynamics described by Eq.~\eqref{eq:hamiltonian5} comprises both the resonant Rabi oscillation of the electronic transition $\ket{0}_e \leftrightarrow \ket{1}_e$ in the $\ket{1}_n$ subspace of the nuclear qubit and the off-resonant Rabi oscillation of the detuned $\ket{0}_n$ subspace with the Rabi frequency 
\begin{equation}
\Omega=\frac{1}{2}\sqrt{B_1^2+\left( A_{||}^{15\rm N} \right)}
\label{eq:synccondition}
\end{equation}
around an axis in the $x$-$z$ plane determined by the driving strength $B_1$ and the hyperfine coupling $A_{||}^{15\rm N}$.

Both on and off-resonant Rabi oscillations during the gate time $t_g$ are described by the unitary operator $U=e^{-i\Tilde{H}t_g}$. Choosing $t_g=\pi / B_1$ leads to a $\pi$-rotation of the electron spin around the $x$-axis in the nuclear spin $\ket{1}_n$-subspace. In order to compensate a relative phase between the nuclear spin subspaces, we supplement the driven operation to form the actual gate operation $U_{\rm act}=R_z^n(\pi/2) e^{-it\Tilde{H}}$ where the rotation $R_z^n(\theta)=e^{-i\theta I_z^{15\rm N}}$ of the nuclear spin around the $z$-axis by an angle $\theta$ may be executed virtually. 

As may be seen from Eq.~\eqref{eq:hamiltonian5}, the hyperfine interaction between the electron and nuclear spin acts as a native nuclear spin controlled $z$-rotation on the electronic qubit. This interaction is always on, resulting in rotation of the nuclear spins dependent on the respective electron spin projection, thus the system natively performs a CZ$(\theta)$ gate with $\theta=-iA_{||}^{15\rm N}t$ during the duration of the electron spin control. Consequently, the actual gate $U_{\rm act}$ differs from a C$_n$NOT$_e$ where the error probability depends on the phase $\phi=A_{||}^{15\rm N}t_g$ of the $z$ rotation at the gate time $t_g$. 

While this interaction may be exploited for nuclear spin control \cite{Taminiau2012,Kolkowitz2012,Zhao2012_2,Bradley2019}, in the case of resonant electron spin control, it further reduces the fidelity of the conditioned electron spin rotation. Thus, we introduce a waiting time $t_w$ into the gate scheme. The Hamiltonian of the free evolution during the waiting time in the interaction picture simply reads as
\begin{equation}
    \Tilde{H}_0=
    \begin{pmatrix}
        0 &  0 & 0 & 0\\
         0 & -A_{||}^{15\rm N} & 0 & 0 \\
       0  & 0 & 0 & 0\\
        0 & 0  & 0 & 0
    \end{pmatrix},
\end{equation}
with the corresponding unitary evolution for the waiting time $t_w$ given by $U_0 (t=t_w)=e^{-i\Tilde{H}_0 t_w}$.
The waiting time $t_w$ is chosen such that in the absence of a driving field, $B_1=0$, the acquired phase is compensated, which is achieved by 
\begin{align}
    t_w=2\pi/A_{||}^{15\rm N}-t_g,
    \label{eq:waitingtime}
\end{align}
resulting in the gate operation $U_{\rm act,CNOT}(t_g+t_w)=U_0(t_w)  U_{act}$.
We calculate the average gate fidelity \cite{Pedersen2007} of this operation according to 
\begin{equation}
    F_{\rm av}
    =\frac{d+\left| \textup{Tr}\left(U^\dagger_{\rm CNOT}U_{\rm act,CNOT}\right) \right|^2}{d(d+1)},
    \label{eq:gatefidelity}
\end{equation}
where $d=4$ denotes the dimension of the Hilbert space and the ideal evolution is the CNOT gate,
\begin{equation}
U_{\rm CNOT}=\begin{pmatrix}
    1 & 0 & 0 & 0 \\
    0 & 1 & 0 & 0 \\
    0 & 0 & 0 & 1 \\
    0 & 0 & 1 & 0 \\
\end{pmatrix}.
\end{equation}

\begin{figure}
    \centering
\includegraphics[width=0.95\linewidth]{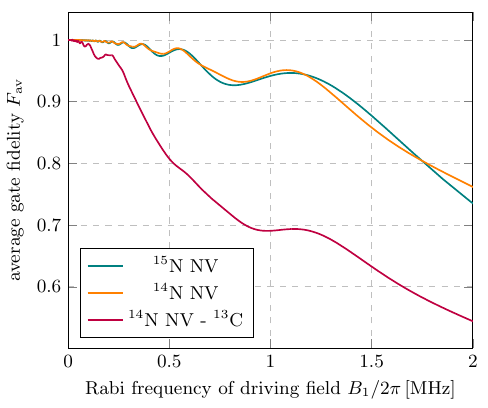}
    \caption{Average gate fidelity $F_{\rm av}$ of a C$_n$NOT$_e$-gate in an NV center as function of the Rabi frequency of the driving field $B_1$, as given by Eq.~\eqref{eq:gatefidelity}. Differently colored lines correspond to different couplings between the electron spin of the NV center and nuclear spins of different nitrogen isotopes and carbon-13 atoms.}
    \label{fig:Fidelity1}
\end{figure}

In Fig.~\ref{fig:Fidelity1}, the average gate fidelity $F_{\rm av}$ is plotted as a function of the Rabi frequency of the driving field $B_1$.

In the regime where the detuning is large compared to the strength of the driving field $A_{||}^{15\rm N} \gg B_1$, the $\ket{0}_n$ subspace is barely affected and as a consequence, the correction induced by $U_0(t_w)$ compensates the error induced by the hyperfine interaction. However, for faster gate operations, the ratio $B_1 / A_{||}^{15\rm N}$ increases, and hence the probability to induce an unwanted electron spin-flip in the $\ket{0}_n$ nuclear spin subspace also grows and leads to a reduction of the gate fidelity. With this, the rotation axis in the Bloch sphere tilts towards the $x$-axis and thus, the correction with $U_0$ increasingly fails and depending on the value of $B_1$ even enhances the error rate. In the regime of strong driving, $A_{||}^{15\rm N} \ll B_1$, the fidelity reaches its minimum $F_{\rm av}=\frac{2}{5}$ as the detuning is vanishing against the driving strength and the rate of the unwanted electronic spin flips equals the rate of the driven transition.

\begin{figure}
    \centering
    \includegraphics[width=\linewidth]{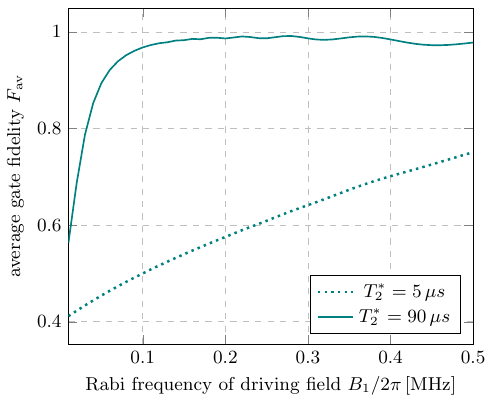}
    \caption{Average gate fidelity $F_{\rm av}$ in the presence of noise due to spin impurities for different values of $T_2^*$. Clearly, the fidelity of operations in the weak driving regime is reduced significantly in both cases.}
    \label{fig:plotnoise}
\end{figure}

One way to suppress unwanted spin-flips is thus to operate in the weak-driving regime $A_{||}^{15\rm N} \gg B_1$, resulting in long gate times. To capture this impact on the overall fidelity of the gate, we include the relaxation process of the central electron spin due to the nuclear spin bath in our calculations (see Sec.~\ref{sec:noisemodel}). In Fig.~\ref{fig:plotnoise}, we plot the resulting average gate fidelity $F_{\rm av}$ in dependence of the driving strength $B_1$ for different values of $T_2^*$. Clearly, operations in the weak driving regime lead to a significant loss of fidelity in all cases.

To overcome this limitation in gate speed, current state-of-the art gate schemes for conditioned rotations of the electron spin rely heavily on complicated, digitally designed pulse sequences based on optimal control algorithms \cite{Dolde2014,Waldherr2014,Rong2015,Xie2023}. 

In contrast to that approach, we propose a different strategy which suppresses unwanted spin flips not by avoiding off-resonant excitations but by choosing the free parameters in the gate scheme such that synchronization cancels out erroneous effects. By doing so, we make use of similar schemes which have been proposed theoretically \cite{Russ2018,Heinz2021} and demonstrated experimentally \cite{Noiri2022} to achieve high-fidelity two-qubit gate operations between spin qubits in semiconductor quantum dots. These approaches employ a synchronization of the Rabi frequencies of both the resonant and the off-resonant transitions such that the Bloch-vector of the off-resonant transition will perform a $2\pi$ rotation around an axis in the $x$-$z$ plane during the gate time $t_g$. This is the case whenever the synchronization condition,
\begin{equation}
    B_1 = \frac{2n+1}{m} \Omega, 
\end{equation}
with integers $n\ge 0$ and $m>0$, is fulfilled.

In case of the system at hand, this condition is fulfilled for certain ratio between the driving field $B_1$ and the hyperfine interaction $A_{||}^{15\rm N}$ which obey
\begin{equation}
    B_1^{\rm sync}(n,m)=A_{||}^{15\rm N}\sqrt{\frac{(2n+1)^2}{4m^2-(2n+1)^2}}.
    \label{eq:sync}
\end{equation}
Due to the synchronization effect for values of the driving strength where $B_1=B_1^{\rm sync}$, the free evolution of the system is compensated, and the operation is equal to a C$_n$NOT$_e$ up to single qubit gates which correct relative phases between both nuclear subspaces according to 
\begin{equation}
U_{\rm act,CNOT}^{sync}=R_z^n(\theta) U_{\rm act},
\end{equation}
where
\begin{equation}
\theta =
\begin{cases}
    \pi/2+A_{||}t_g/4, & \text{if}\: m \:\text{even},\\
    \pi/2+A_{||}t_g/4+\pi, & \text{if} \: m \:\text{odd}.
\end{cases}
\end{equation}

\begin{figure}
    \centering
\includegraphics[width=0.95\linewidth]{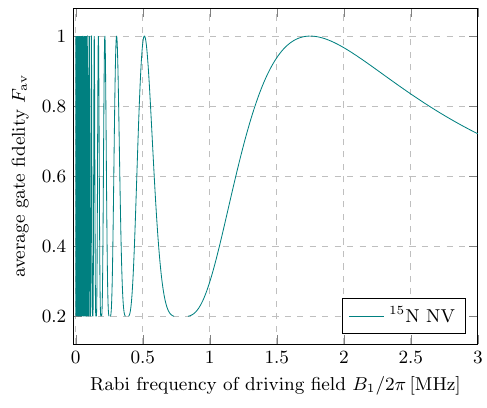}
    \caption{Synchronization effect for the CNOT gate. The average gate fidelity $F_{\rm av}\left(U_{\rm CNOT},U_{\rm act}\right)$ of the CNOT gate is plotted as a function of the Rabi frequency of the driving field $B_1$. The values $B_1^{\rm sync}$ where the resonantly driven and the off-resonant transition are synchronized are visible where the fidelity reaches its maximum $F_{\rm av}=1$. In contrast to Fig.~\ref{fig:Fidelity1}, no  CZ gate is performed to correct the free evolution of the system.  For stronger driving regimes, no synchronization between the resonant and off-resonant driving is possible and the fidelity steadily decreases until it reaches its lowest value of $F_{\rm av}=2/5$ (not shown in the plot). For weak driving, the fidelity oscillates between its maximum and minimum values while the period of this oscillation increases steadily as the free evolution of the system becomes increasingly faster relative to the driving strength.}
   \label{fig:Fidelity2}
\end{figure}

\begin{figure*}
    \centering
\includegraphics[width=0.8\textwidth]{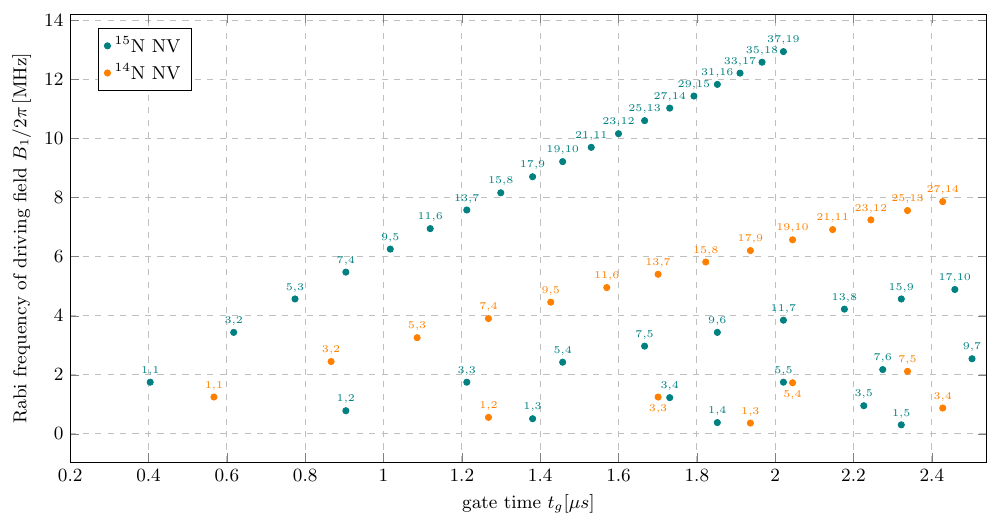}
    \caption{Driving strength $B_1$[MHz] as a function of the gate time $t_g$[$\mu\,\text{s}$] for several choices of $B_1^{\rm sync}$. Differently colored dots correspond to couplings between the electron spin of the NV center and nuclear spins of different nitrogen isotopes. Each dot is labeled with the respective values for $n,m$ defined by Eq.~\eqref{eq:sync}.}
    \label{fig:syncpoints}
\end{figure*}

In Fig.~\ref{fig:Fidelity2}, we plot the average gate fidelity $F_{\rm av}$ of this operation for odd values of $m$ in Eq.~\eqref{eq:sync}. The values of $B_1^{\rm sync}$ where Eq.~\eqref{eq:sync} is fulfilled are visible as points where the fidelity reaches its maximal value, $F_{\rm av}=1$. Choosing $m=1$ and $n=0$ for the fastest gate operation yields the maximum value for driving strength $B_1^{\rm sync}(n=0,m=1)=A_{||}^{15\rm N}/\sqrt{3}\approx 1.75\,\textup{MHz}$ and $\Tilde{B}_1 \approx 2.47\,\textup{MHz}$ for which the synchronization conditions are fulfilled, allowing for a gate time of $t_g\approx 0.4\,\mu \text{s}$. Other values of $B_1^{\rm sync}$ will lead to slightly longer gate times, as Fig.~\ref{fig:syncpoints} shows. For larger values of $B_1$, no synchronization between the resonant and off-resonant driving is possible and the fidelity steadily decreases until it reaches its lowest value of $\nicefrac{2}{5}$. For smaller values of $B_1$, the fidelity oscillates with an increasing period as the free evolution of the system becomes increasingly faster relative to the driving strength. 
\begin{figure}
    \centering
    \includegraphics[width=\linewidth]{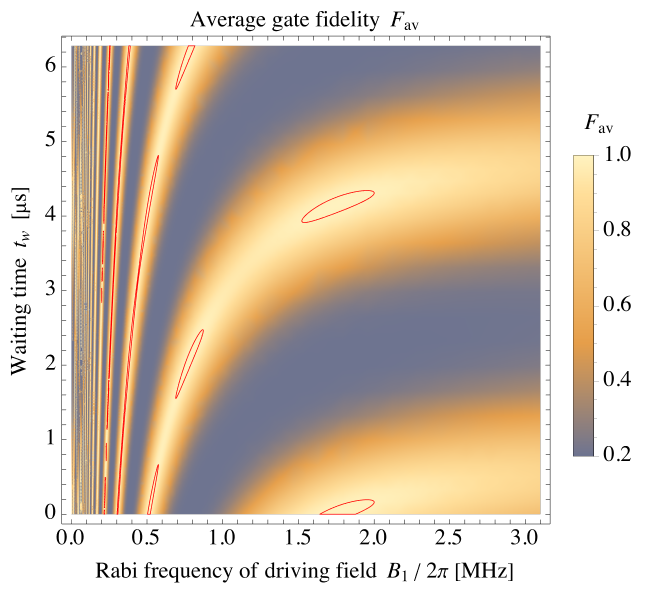}
    \caption{Average gate fidelity $F_{\rm av}$ as a function of driving strength $B_1$[MHz] and the waiting time $t_w$. Marked regions are those where $F_{\rm av}>0.99$.}
    \label{fig:fidelity}
\end{figure}
For other values of $B_1$ where the two Rabi oscillations are not fully synchronized, $F_{\rm av}$ can be enhanced by making use of the system's native CZ$(\theta=A_{||}^{15\rm N}t_w)$ operation. By doing so, the range of driving strengths allowing for high-fidelity operations is increased significantly. Fig.~\ref{fig:fidelity} shows the average gate fidelity for different values of $B_1$ and $t_w$. Clearly, in some regions where the Rabi-oscillations are not fully synchronized, an additional rotation around the $z$-axis corrects the main part of the error induced by the off-resonant driving. Regions marked in red in Fig.~\ref{fig:fidelity} are those where $F_{\rm av} \in [0.99,1]$, thus high-fidelity operations may be achieved. Clearly, this significantly extends the range of possible choices for the driving strength. 

\section{Synchronisation protocol for a Qubit-Qutrit gate operation}
\label{sec2}
Most experiments demonstrating the use of an NV center as a quantum register build on systems where the nitrogen atom is not of the implanted $^{15}$N isotope type but consists of the $^{14}$N atom occurring with a much higher natural concentration, with the nuclear spin $I^{14\rm N}=1$. In this section, we show how the synchronization protocol may be extended to this type of system.
Again, Eqs.~\eqref{eq:totalhamiltonian}--\eqref{eq:Hamiltonian3} define the model, with one coupled nuclear spin and thus $j=14N$ with $I^{14\rm N}=1$. The hyperfine coupling and the gyromagnetic ratio are given by $\gamma_n^{14\rm N}/(2\pi)=3.1\,\text{MHz/T}$, $A_{||}^{14\rm N}/(2\pi)=-2.16\,\text{MHz}$, and $A_\perp^{14\rm N}/(2\pi)=-2.7\,\text{MHz}$ \cite{Zaiser2019}. Additionally, the system exhibits a quadrupole splitting with $Q^{14\rm N}/(2\pi)=-4.96\,\text{MHz}$ such that Eq.~\eqref{eq:Hamiltonian2} reads as
\begin{equation}
    H_n^{{14N}} = \gamma_n^{14\rm N} B_z I_z^{14\rm N} + Q^{14\rm N} (I^{14\rm N}_z)^2.
\end{equation}

\begin{figure}[h]
    \centering
    \includegraphics[width=\linewidth]{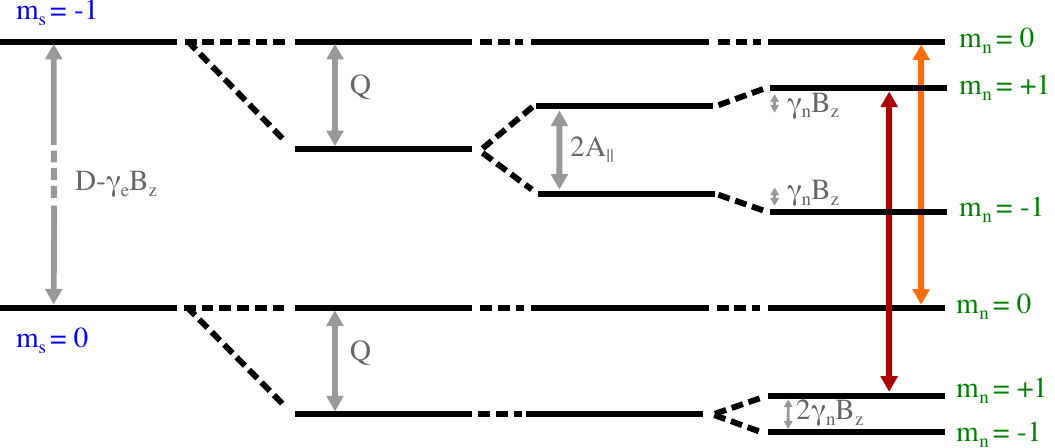}
    \caption{Energy level scheme of a NV center with a $^{14}$N atom. Only the electronic computational levels $m_s=0$, $m_s=-1$ are displayed, with the $m_s=1$-transition being far detuned. The red and orange arrows indicate the transitions under study.}
    \label{fig:energylevels14N}
\end{figure}

Again, we define the electronic qubit states to be $\ket{m_s=0}\equiv\ket{0}_e$ and $\ket{m_s=-1}\equiv\ket{1}_e$, assume that the driving strength is constant, according to Eq.~\eqref{eq:drive}, and that the  $m_s=+1$ level is far off-resonant.  In this case, the driving strength $B_1$ is small compared to the detuning $\delta$ (cf. Eq.~\eqref{eq:delta}). The computational subspace of the nuclear spin is defined by $\ket{m_n=0}\equiv\ket{0}_{\rm N}$ and $\ket{m_n=1}\equiv\ket{1}_{\rm N}$. In case of the nuclear spin triplet, the detuning of the $\ket{m_n=-1}$ level is of the same order of magnitude as the driven transition and thus has to be included as a leakage space into the calculations. The energy level scheme of the system is depicted in Fig.~\ref{fig:energylevels14N}. 
We transform the Hamiltonian into the interaction picture of the free energies of the system, and then transform into the rotating frame of the drive. The unitary transformation for these two steps is given by $U=e^{it\omega S_z+it[Q^{14\rm N}(I_z^{14\rm N})^2+\gamma_n B_z I_z^{14\rm N}]}$. In order to achieve a C$_n$NOT$_e$, we address the transition $\ket{1}_{\rm N}\ket{0}_e \leftrightarrow \ket{1}_{\rm N}\ket{1}_e$ with the frequency $\omega=D-B_z\gamma_e -A_{||}^{14\rm N}$, which is indicated by the red arrow in Fig.~\ref{fig:energylevels14N}. With this, the Hamiltonian of the electronic computational subspace can be written in the 6x6 matrix form,
\begin{equation}
    \Tilde{H}=
    \begin{pmatrix}
        0 &  \nicefrac{B_1}{2} & 0 & 0 & 0 & 0\\
         \nicefrac{B_1}{2} & 0 & 0 & 0 & 0 & 0\\
       0  & 0 & 0 & \nicefrac{B_1}{2} & 0 & 0\\
        0 & 0  & \nicefrac{B_1}{2} & -A_{||}^{14\rm N} & 0 & 0\\
        0 & 0 & 0 & 0 & 0 &  \nicefrac{B_1}{2}\\
        0 & 0 & 0 & 0 & \nicefrac{B_1}{2} & -2A_{||}^{14\rm N}
    \end{pmatrix}, 
    \label{eq:hamiltonian6}
\end{equation}
with the basis $\ket{11}, \ket{10}, \ket{1-1}, \ket{01}, \ket{00}, \ket{0-1},\ket{-11}, \ket{-10}, \ket{-1-1}$ where $\ket{ij}=\ket{i}_{\rm N} \ket{j}_e $
and we have dropped fast oscillating terms within a RWA. As before, the  unitary time evolution generated by this Hamiltonian is given by $U_0=e^{-i\Tilde{H}t}$.
Analogous to the case of the 4-level system, the corresponding gate fidelity $F_{\rm av}$ with $U_{\rm act}=U_0(t_w) e^{-i\Tilde{H}t_g}$ depends strongly on the ratio between $A_{||}^{14\rm N}$ and $B_1$ (see orange line in Fig.~\ref{fig:Fidelity1}).
\begin{figure}
    \centering
    \includegraphics[width=0.95\linewidth]{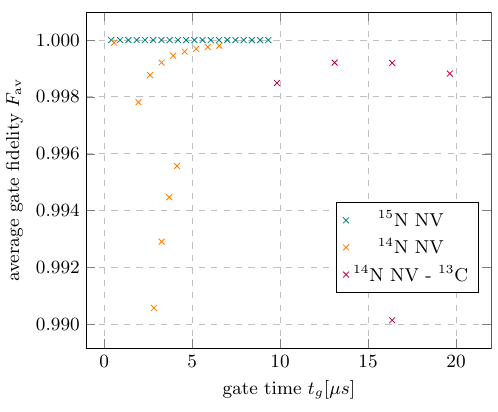}
    \caption{Fidelity of synchronized gates corresponding to values $B_1^{\rm sync}$ of the drive strength, as a function of the gate time $t_g\,$[$\mu$s]. Differently colored lines correspond to an NV center coupling to nuclear spins of different isotopes.}
    \label{fig:syncpoints14N}
\end{figure}

\begin{figure}[h]
    \centering
    \includegraphics[width=\linewidth]{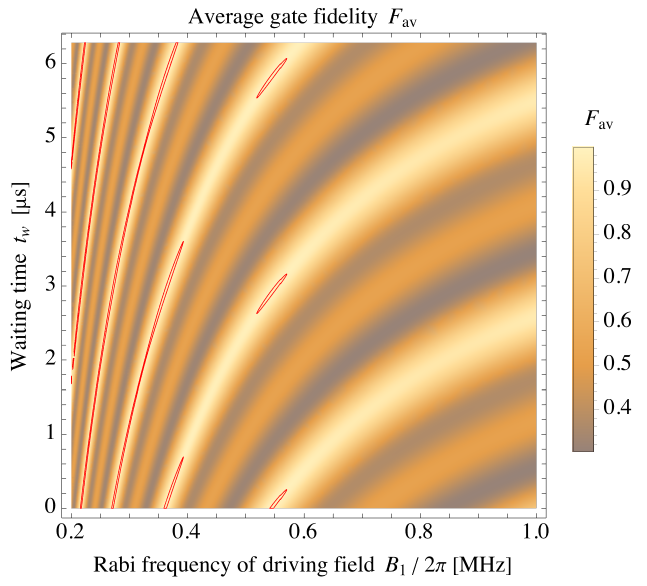}
    \caption{Average gate fidelity $F_{\rm av}$ of a C$_n$NOT$_e$ gate on an NV center comprising a $^{14}$N atom as a function of driving strength $B_1$ and the waiting time $t_w$.  Regions marked in red are those where $F_{\rm av}>0.99$.}
    \label{fig:fidelity6}
\end{figure}

Contrary to the case of the NV-$^{15}$N system where the desired C$_n$NOT$_e$-operation has to be synchronized with just one off-resonant transition, in case of the NV-$^{14}$N system, two unwanted transitions are addressed off-resonantly by the drive (Eq.~\eqref{eq:hamiltonian6}). Achieving synchronization such that both off-resonant transitions will have a vanishing probability for unwanted spin-flips leads to diophantine equations and Boolean conditions for three integers for which no analytical solution for $F_{\rm av}=1$ exists. 
However, some solutions $B_1^{\rm sync}$ still lead to high-fidelity operations with $F_{\rm av}(B_1^{\rm sync})>0.99$ and allow for fast operation times (see Fig.~\ref{fig:syncpoints14N}). Analogous to the case of the NV center comprising an $^{15}$N atom, the range of the driving field strength leading to a fidelity above $0.99$ may be extended by making use of the  native CZ gate of the system, cf. Fig.~\ref{fig:fidelity6}.

A gate operation that is locally equivalent (up to single-qubit gates) to a C$_n$NOT$_e$ is achieved by driving the transition $\ket{m_n=0,m_s=0} \leftrightarrow \ket{m_n=0,m_s=1}$ which is indicated by the orange arrow in Fig.~\ref{fig:energylevels14N}. In this case, due to the inherent symmetry of the system, the same condition as in case of the NV-$^{15}$N system holds, cf.\ Eq.~\eqref{eq:synccondition}, with a vanishing gate error with respect to the off-resonant drive. Here, the fastest gate is achieved within $0.56\,\mu\text{s}$ (see orange points in Fig.~\ref{fig:syncpoints}). This is a similar order of magnitude as gate times achieved with optimal control sequences for this system, which are of the duration $\sim 0.6-1.5\,\mu s$ \cite{Rong2015,Xie2023}.

\section{Entangling gate conditioned on states of a multi-level system}
\label{sec3}
In this section, we extend the synchronization scheme to include $^{13}$C atoms which couple to the NV center via the hyperfine interaction. The free evolution of an isolated $^{13}$C nuclear spin in an external magnetic field is described as $H_n^{13\rm C}=B_z \gamma_n^{13\rm C} I_z^{13\rm C} $ with $\gamma_n^{13\rm C}/(2\pi)=10.705\,\text{MHz}$ \cite{Zaiser2019}. $^{13}$C nuclei are spin doublets, $I^{13\rm C}=1/2$, thus the computational subspace of the isolated system is defined as $\ket{m_n=1/2}\equiv\ket{0}_{\rm C}$ and $\ket{m_C=-1/2}\equiv\ket{1}_{\rm C}$.
The equations of motion of the composite system read as
\begin{equation}
    H= H_e + H_n^{14\rm N} + H_n^{13\rm C} + H_i^{14\rm N}+H_i^{13\rm C}+H_d,
    \label{eq:multilevelsystem}
\end{equation}
where the interaction part between nuclear and electronic spins is described by $H_i^{13\rm C}=\textbf{\textit{SA}}^{13\rm C}\textbf{\textit{I}}^{13\rm C}$. Based on the secular approximation \cite{Dutt2007}, $\textbf{\textit{A}}^{13\rm C}$ contains two non-zero matrix elements $A^{13\rm C}_{zz}$ and $A^{13\rm C}_{zx}$. Their values depend on the position and relative axis of the respective $^{13}$C isotope towards the main axis of the NV center. Here, we assume a choice of the coupled $^{13}$C spins such that $A^{13\rm C}_{zx} \ll A^{13\rm C}_{zz}$ which allows us to neglect the influence of the perpendicular component of the hyperfine tensor, thus to assume $A^{13\rm C}_{zx} \approx 0$. Note that when the NV center is coupled to several  carbon-13 nuclear spins, individual spins may be distinguished by their individual hyperfine component $A^{13\rm C}_{zz}$.

Including these considerations, we transform the Hamiltonian in the interaction picture of the free nuclear energies and in the rotating frame of the driving frequency with the unitary transformation 
$U=e^{it\omega S_z+it(H_n^{{14N}} + H_n^{13\rm C})}$ and drop fast oscillating terms. With this, Eq.~\eqref{eq:multilevelsystem} becomes
\begin{equation}
    \Tilde{H} = \Tilde{H}_0 + B_1/2,
    \label{eq:hamiltonian7}
\end{equation}
where 
\begin{align}
  \Tilde{H}_0=&(DS_z^2+ \gamma_{e}B_z - \omega) S_z \\
  &+A^{{14N}}_{||}S_z I^{14\rm N}_z + A_{zz}^C S_z I^{13\rm C}_z.
\end{align}
When neglecting the $m_s=1$-subspace, the coupled system is reduced to twelve non-degenerate energy levels.  

Due to the hyperfine coupling of the nuclear and electron spin, the electronic spin of the $^{14}\text{N-}^{13}$C-NV system may be addressed conditioned in principle on any product state of the two nuclear spins, resulting in six different, frequency-selective transitions. 

Exemplary, we consider the $\ket{0}_{\rm C} \ket{1}_{\rm N}\ket{0}_e\leftrightarrow \ket{0}_{\rm C}\ket{1}_{\rm N}\ket{1}_e$ transition with the frequency of the driving field set to $\omega = D-\gamma_{e}B_z+A^{14\rm N}_{||}+A^{13\rm C}_{zz}/2$. This results in one resonant driven transition and five off-resonantly addressed transitions resulting in unwanted spin-flips.

Equivalently to the cases discussed above, choosing the gate time accordingly results in a CC$_n$NOT$_e$ operation where the electronic spin is flipped conditioned on the state of two nuclear spins. This is locally equivalent to a Toffoli gate. As explained in Sec.~\ref{sec1}, the free evolution is corrected by applying $U_0(t_w)=e^{-i\Tilde{H}_0t_w}$ for the time $t_w=\nicefrac{2\pi}(A^{14\rm N}_{||}+A^{13\rm C}_{zz}/2)-t_g$.

Similarly to the previously discussed smaller systems, the average gate fidelity of the operation strongly deteriorates with increasing driving strength (see purple line in Fig.~\ref{fig:Fidelity1}). As an example, we set $A^{13\rm C}_{zz}=0.43\,\text{MHz}$, and thus assume a strong coupling between the NV center and the carbon isotope \cite{Zaiser2019}. 

While no analytical solution for $B_1$ exists which fully synchronizes all six transitions given by Eq.~\eqref{eq:hamiltonian7}, values for $B_1$ may be found where $F_{\rm av} \geq 0.99$ allowing a significant reduction of the error while achieving fast gate operations due to a significantly higher driving strength. Fig.~\ref{fig:syncpoints14N} depicts the gate fidelity for selected values of the driving strength, again plugging $A^{13\rm C}_{zz}=0.43\,\text{MHz}$ into our calculations. Note that the gate time demonstrated with optimal control is $t_g\approx32\,\mu\text{s}$ \cite{Waldherr2014} while for the system at hand, we predict the lowest value as $t_g\approx10\,\mu\text{s}$.

\section{Conclusion}
\label{sec:conclusion}
In this work, we have proposed fast schemes for entangling gates of high fidelity on a nitrogen-vacancy center, where the electron spin is flipped conditioned on the state of one or several nuclear spins, either only of the intrinsic nitrogen atom or also including a nuclear spin of the surrounding carbon-13 atoms in the diamond lattice. These schemes eliminate the effects of power broadening in the ODMR spectrum - thus of unwanted spin flips due to off-resonantly driving additional transition lines - by exploiting synchronization effects between all relevant transitions in the respective systems, allowing for fast gate operations significantly shorter than $1\,\mu\text{s}$. Also, these schemes allow to drive the system in the strong regime when taking slightly longer gate times into account. In addition to that, by making use of the  native CZ generated by the hyperfine interaction, the range of driving strength leading to a vanishing error rate may be extended significantly. We have predicted vanishing error rates for a two-qubit C$_n$NOT$_e$ gate on a qubit-qubit system and  a qubit-qutrit system as well as fidelities above $>0.99$ for a Toffoli CC$_n$NOT$_e$ operation on a multi-level system. Our results contribute to the goal of high-fidelity control with error rates $<1\%$ for the perspective of fault-tolerant quantum computation with defect-based systems.
Further work could investigate extending the protocol to different types of single defects in solids such as silicon vacancy centers in diamond, and the adaption of these protocols in state-of-the art DDrf schemes for NV centers, where nuclear spins are driven conditioned on the electron spin state.

\section*{Acknowledgements}
We thank Vadim Vorobyov and Dzhavid Dzhavadzade for fruitful discussions. We acknowledge funding from the state of Baden-Württemberg through the Kompetenzzentrum Quantum Computing, project QC4BW.

\begin{appendix}
\section{Noise Model}
\label{sec:noisemodel}
To capture the impact of long gate times in the weak driving regime on the overall fidelity of the gate, we include the relaxation process of the central electron spin in our calculations \cite{Liu2012}. The interaction of the NV center with its environment may be described as a central spin decoherence problem, where the central electron spin couples to an anisotropic hyperfine field formed by the surrounding $^{13}C$ nuclear spins, while interaction between the nuclear spins is small against this coupling and may thus be neglected. This leads to strong quantum fluctuations if the strength of the DC magnetic field is such that the Zeeman energy $\gamma_CB_z$ of the $^{13}C$ nuclear spins is comparable to their hyperfine couplings $\textbf{A}^{13\rm C}$ - thus not too strong nor too weak. The fluctuations are captured by the local Overhauser bath operator $\delta B$ where we put $\delta B_x=\delta B_y\equiv 0$ as the zero field splitting is in the GHz-range and thus much larger than the hyperfine coupling with the carbon-13 atoms. With this, Eq.~\eqref{eq:Hamiltonian1} is altered to \cite{Maze2008,Zhao2012_2}
\begin{equation}
    H_e(\delta B_z)=\textit{\textbf{SDS}}+(B_z \gamma_{e}+\delta B_z) S_z .
    \label{eq:overhauserfield}
\end{equation}

Assuming a large number of unpolarised nuclear bath spins, the local Overhauser field is captured with a zero-mean Gaussian distribution with a width is given by $\sigma=\sqrt{\langle \delta B_z^2 \rangle-\langle \delta B_z \rangle^2}$. As the timescale of the fluctuations is large against the gate time, we calculate the average gate fidelity of the noise process as 
\begin{equation}
    \Bar{F}_{\rm av}(\sigma)=\int_{-\infty}^{\infty} F_{\rm av}(\delta B) p(\delta B) \text{d} \delta B,
\end{equation}
with 
\begin{equation}
p(\delta B)= 1/\sqrt{2 \pi \sigma^2} \exp{\left\{ -1/2 (\delta B/\sigma)^2 \right\}},
\end{equation}
and $F_{\rm av}$ the gate fidelity (Eq.~\eqref{eq:gatefidelity}) where the free evolution of the electron spin is described by Eq.~\eqref{eq:overhauserfield}.

The electron spin decoherence time $T_2^*$ depends on the ambient temperature, the external magnetic field, and the isotope density in the surrounding material. Reported values are $2-7\,\mu {\rm s}$ \cite{Liu2012,Bradley2019} for material with a natural $^{13}$C abundance of 1.1\% and up to $T_2^*\approx 90\,\mu {\rm s}$ for isotopically purified material \cite{Ishikawa2012,Fang2013}. Due to the slow nuclear spin bath dynamics, the electronic $T_2$ time may be extended up to more than $1\,{\rm s}$ by the use of dynamical decoupling sequences \cite{Abobeih2018}. This technique is applied for nuclear spin control. Recent work has successfully incorporated the time-dependent nature of the spin bath into the electron control pulse design \cite{Xie2023}.

\end{appendix}

\bibliography{bibliography.bib}
	
\end{document}